
\input harvmac

\input epsf

\newcount\xrefpos \xrefpos=0
\newcount\yrefpos \yrefpos=0
\newcount\xput \xput=0
\newcount\yput \yput=0
\def\refpos#1 #2 #3{\global\xrefpos=#1 \global\yrefpos=#2
                         \rlap{$\smash{#3}$}}
\def\put #1 #2 #3{\xput=#1 \yput=#2
                  \advance\xput by -\xrefpos
                  \advance\yput by -\yrefpos
                  \rlap{\kern\the\xput truebp
                       \vbox to 0pt{\vss\hbox{\footnotefont$\displaystyle  #3$}
                        \kern\the\yput truebp}}}
\def\beginlabels\refpos#1\endlabels{\hbox{$\refpos#1$}}


\def\figin{\epsfcheck\figin}\def\figins{\epsfcheck\figins}
\def\epsfcheck{\ifx\epsfbox\UnDeFiNeD
\message{(NO epsf.tex, FIGURES WILL BE IGNORED)}
\gdef\figin##1{\vskip2in}\gdef\figins##1{\hskip.5in}
\else\message{(FIGURES WILL BE INCLUDED)}%
\gdef\figin##1{##1}\gdef\figins##1{##1}\fi}
\def\DefWarn#1{}
\def\figinsert{\goodbreak\midinsert}
\def\ifig#1#2#3{\DefWarn#1\xdef#1{fig.~\the\figno}
\writedef{#1\leftbracket fig.\noexpand~\the\figno}%
\figinsert\figin{\centerline{#3}}\bigskip\centerline{\vbox{\baselineskip12pt
\advance\hsize by -1truein\noindent\footnotefont{\bf Fig.~\the\figno:} #2}}
\bigskip\endinsert\global\advance\figno by1}

\font\cmss=cmss10 \font\cmsss=cmss10 at 7pt
\def\IR{\relax{\rm I\kern-.18em R}}
\def\IZ{\relax\ifmmode\mathchoice
{\hbox{\cmss Z\kern-.4em Z}}{\hbox{\cmss Z\kern-.4em Z}}
{\lower.9pt\hbox{\cmsss Z\kern-.4em Z}}
{\lower1.2pt\hbox{\cmsss Z\kern-.4em Z}}\else{\cmss Z\kern-.4em Z}\fi}
\def\Im{\mathop{\rm Im}}
\def\Re{\mathop{\rm Re}}

\def\rpsi{\psi}
\def\rpsib{\underline{\psi}}
\def\rxi{\xi}
\def\rxib{\underline{\xi}}
\def\rh{h}
\def\rhb{\underline{h}}
\def\re{e}

\def\zp{z_{\scriptscriptstyle +}}

\lref\chop{ M.W.~Choptuik,
``Universality and Scaling in Gravitational Collapse of a Massless
Scalar Field,''
{\it Phys.\ Rev. Lett.}\ {\bf 70} (1993) 9.}
\lref\abev{A.M.~Abrahams and C.R.~Evans,
``Critical Behavior and Scaling in Vacuum Axisymmetric Gravitational
Collapse,''
{\it Phys\ Rev.\ Lett.}\ {\bf 70} (1993) 2980;
{\it Phys.\ Rev.}\ {\bf D49} (1994) 3998.}
\lref\evco{C.R.~Evans and J.S.~Coleman,
``Critical Phenomena and Self-Similarity in the Gravitational
Collapse of Radiation Fluid,''
{\it Phys.\ Rev.\ Lett.}\ {\bf72} (1994) 1782, gr-qc/9402041.}
\lref\kha{T.~Koike, T.~Hara, and S.~Adachi,
``Critical Behavior in Gravitational Collapse of Radiation Fluid:
A Renormalization Group (Linear Perturbation) Analysis,''
{\it Phys.\ Rev.\ Lett.}\ {\bf 74} (1995) 484,
gr-qc/9503007.}
\lref\mai{D.~Maison,
``Non-Universality of Critical Behaviour in Spherically Symmetric
Gravitational Collapse,''
gr-qc/9504008.}
\lref\hieaa{E.W.~Hirschmann and D.M.~Eardley,
``Universal Scaling and Echoing in Gravitational Collapse of a
Complex Scalar Field,''
{\it Phys.\ Rev.}\ {\bf D51} (1995) 4198, gr-qc/9412066.}
\lref\hieab{E.W.~Hirschmann and D.M.~Eardley,
``Critical Exponents and Stability at the Black Hole Threshold for a
Complex Scalar Field,''
gr-qc/9506078.}
\lref\ehh{D.M.~Eardley, E.W.~Hirschmann, and J.H.~Horne,
``$S$-Duality at the Black Hole Threshold in Gravitational Collapse,''
gr-qc/9505041.}
\lref\gun{C.~Gundlach,
``The Choptuik spacetime as an eigenvalue problem,''
gr-qc/9507054.}
\lref\hast{R.S.~Hamad\'e and J.M.~Stewart,
``The spherically symmetric collapse of a massless scalar field,''
gr-qc/9506044.}
\lref\beol{M.J.\ Berger and J.\ Oliger,
``Adaptive Mesh Refinement for Hyperbolic Partial Differential
Equations,''
{\it J.\ Comput.\ Phys.} {\bf 53} (1984) 484.}

\Title{\vbox{\baselineskip12pt\hbox{DAMTP/R-95/54}
\hbox{gr-qc/9511024}
}}
{\vbox{
\centerline{Continuous Self-Similarity and $S$-Duality}
}}

\centerline{
{Rufus S.\ Hamad\'e}\footnote{$^\dagger$}{Email address:
rsh1000@damtp.cam.ac.uk},
{James H.\ Horne}\footnote{$^\ddagger$}{Email address: jhh20@damtp.cam.ac.uk},
and {John M.\ Stewart}\footnote{$^*$}{Email address: john@damtp.cam.ac.uk}}
\vskip .12in
\centerline{\sl Department of Applied Mathematics and Theoretical Physics}
\centerline{\sl	University of Cambridge}
\centerline{\sl	Silver Street}
\centerline{\sl	Cambridge CB3 9EW}
\centerline{\sl	Great Britain}

\bigskip
\centerline{\bf Abstract}
We study the spherically symmetric collapse of the axion/dilaton
system coupled to gravity. We show numerically that the critical
solution at the threshold of black hole formation is continuously
self-similar. Numerical and analytical arguments both demonstrate that
the mass scaling away from criticality has a critical exponent of
$\gamma = 0.264$.

\Date{November 6, 1995}

\newsec{Introduction}

Consider a spherically symmetric spacetime filled by a massless scalar
field $\Psi$. The field $\Psi$ evolves according to
\eqn\epsi{
g^{ab} \nabla_a \nabla_b \Psi = 0 \; ,
}
where the geometry is determined by Einstein's field equations in the
form
\eqn\eein{
R_{ab} = \kappa^2 \partial_a \Psi \partial_b \Psi \;,
}
where $\kappa^2 = 8 \pi G$. At first sight this problem would appear
to be totally predictable and boring. Appearances are however
deceptive.

The system of equations~\epsi\ and \eein\ has been studied
numerically, as a Cauchy problem in~\chop\ and as a characteristic
initial value problem in~\hast. In both cases the initial field can be
specified in an arbitrary manner. Choptuik examined the following
problem in~\chop. Suppose the initial data belong to a one-parameter
set. Suppose further that for small values of the parameter $p$, the
field is weak. One expects the evolution to be similar to the solution
of~\epsi\ on Minkowski spacetime; the field evaporates leaving an
empty spacetime. Suppose that for large values of the parameter $p$,
the initial fields are strong. Assuming some form of cosmic censorship
hypothesis, we might expect a black hole to form around a spacetime
singularity. By continuity there should be a critical parameter
$p_{\rm crit}$ separating the two types of evolution.

Choptuik discovered numerically that nearly critical solutions appear
to exhibit discretely self-similar behavior --- structure appears on
ever finer scales~\chop. (This was not discovered in previous
numerical work because an adaptive grid algorithm is needed to resolve
it.) If this behavior is real then the field and curvature can grow
arbitrarily large close to the axis. Further, Choptuik exhibited
in~\chop\ a scaling law. For supercritical evolutions ($p > p_{\rm
crit}$) a black hole forms and he obtained a numerical law estimating
the black hole mass
\eqn\escal{
M_{\rm bh} = c_i (p - p_{\rm crit})^\gamma \; .
}
Here $c_i$ depends on the nature of the initial data but $\gamma
\approx 0.37$ appears to be independent of the initial data. These
results were confirmed in~\hast, and similar results for other forms
of matter have been established in~\refs{\abev,\evco}.

Discrete self-similarity is hard to treat analytically (see
however~\gun), and so theoreticians have looked for other forms of
matter for which continuous self-similarity is possible.  Assuming
continuous self-similarity for the critical solution reduces the
problem to an ordinary differential equation.  A perfect radiation
fluid coupled to gravity was studied in~\evco, where it was
numerically found that the critical solution is indeed continuously
self-similar. Analytical studies of continuously self-similar
solutions where the scaling symmetry mixes with field symmetries have
been made for the complex scalar field~\hieaa\ and for the
axion/dilaton field~\ehh, but it was not demonstrated that the
continuously self-similar solution is indeed the critical solution.

This paper reports on a numerical study of the axion/dilaton problem
using the numerical techniques of~\hast. In \S{2} we describe the
axion/dilaton problem in detail and \S{3} describes the numerical
procedure. Section~4 reviews the conjectured continuously self-similar
critical solution. The main result of the paper is given in \S{5};
there is a continuously self-similar solution which appears to be the
attractor at criticality. In \S{6} we consider slightly supercritical
solutions and confirm both analytically and numerically the scaling
law~\escal. We estimate
\eqn\eanswer{
\gamma_{\rm analytic}  = 0.2641066 \; , \; \gamma_{\rm numeric} =
0.264 \; ,}
giving excellent agreement between theory and experiment. This
critical exponent is however very different from the value $\gamma
\approx 0.37$ found for a massless scalar field~\chop, black body
radiation~\evco, and gravitational radiation~\abev.

\newsec{The axion/dilaton system}

We use the notation of~\ehh. The matter used in this paper occurs in
the $3+1$-dimensional low-energy effective action of string theory. It
consists of gravity coupled to a dilaton $\phi$ and an axion
$\rho$. The axion arises as the dual of the string three-form ${\bf
H}$, via
\eqn\eaxion{
H_{abc} = {1 \over \kappa} e^{4 \phi}
 \epsilon_{abcd} \partial^d \rho \; .
}
The dilaton and axion can be combined into a single complex field
\eqn\etau{
\tau \equiv 2 \rho + i e^{-2 \phi} \; .
}
(We have chosen to rescale $\rho$ by a factor of 2 from the usual
definition.)  Omitting the (presumably) irrelevant gauge fields, the
effective action is
\eqn\eaction{
I = {1 \over 2 \kappa^2} \int \! d^4 x \sqrt{-g} \left( R
- {1 \over 2} { \partial_a \tau \partial^a \bar\tau \over (\Im
\tau)^2} \right)
\; ,}
where $R$ is the scalar curvature.

At the classical level, the model~\eaction\ has an extra $SL(2,\IR)$
symmetry that involves only $\tau$ and leaves the metric
invariant. The symmetry acts on $\tau$ as
\eqn\esltr{
\tau \rightarrow { a \tau + b \over c \tau + d} \; ,}
where $a,b,c,d \in \IR$ and $ad - bc = 1$. This $SL(2,\IR)$ symmetry
is presumably broken by quantum effects down to an $SL(2,\IZ)$
symmetry, which is the popular $S$-duality conjectured to be a
non-perturbative symmetry of string theory.

The equations of motions arising from~\eaction\ are
\eqna\eeom
$$\eqalignno{
R_{ab} - {1 \over 4 (\Im \tau)^2} (\partial_a \tau \partial_b
\bar{\tau} + \partial_a \bar{\tau} \partial_b \tau) & = 0 \; , &\eeom{a}
\cr
\nabla^a\nabla_a \tau + {i \nabla^a \tau \nabla_a \tau \over \Im \tau}
& = 0 \; . &\eeom{b}
}
$$
We will use two approaches to solve these equations. First, we will
integrate them numerically given specified initial conditions. This
will give us a range of numerical solutions, some ending with black
holes, some ending with flat space. Second, we will follow~\ehh\ and
assume that the exactly critical solution is continuously
self-similar. The spacetime can then be derived analytically. We then
compare the answers, and find extremely good agreement.

\newsec{The Numerical Approach}

The numerical integration is based on the approach described
in~\hast. Throughout this paper, we assume spherical symmetry. As
in~\hast, the coordinate system most suitable to the problem at hand
is a double-null coordinate system. We take the spacetime metric to be
\eqn\emet{
ds^2 = - a^2(u,v)\, du \, dv + r^2(u,v) \, d\Omega^2
.}
Double-null coordinates are not unique. The remaining gauge freedom
consists of redefining $u$ and $v$ by $u \rightarrow f(u)$ and $v
\rightarrow g(v)$, where $f$ and $g$ are monotonic functions. We fix
one of the functions by requiring $u = v$ on the axis where $r(u,v) =
0$. The remaining gauge freedom will be fixed by the initial
conditions. See~\hast\ for further details.

With these coordinates, the equations of motion~\eeom{} become
\eqna\eeomuv
$$\eqalignno{
r \, r_{uv} + r_u r_v + {1 \over 4} a^2 = & \; 0 \; , & \eeomuv{a}\cr
a_{uv}/a - a_u a_v/a^2 + r_{uv}/r + \phi_u \phi_v + e^{4\phi} \rho_u
\rho_v = & \; 0 \; , & \eeomuv{b} \cr
r_{uu} - 2 a_u r_u/a + r \left( {\phi_u}^2 + e^{4\phi} {\rho_u}^2
\right) = & \; 0 \; , & \eeomuv{c} \cr
r_{vv} - 2 a_v r_v/a + r \left( {\phi_v}^2 + e^{4\phi} {\rho_v}^2
\right) = & \; 0 \; , & \eeomuv{d} \cr
r \, \phi_{uv} + r_u \phi_v + r_v \phi_u - 2 r \, e^{4 \phi} \rho_u
\rho_v = & \; 0 \; , & \eeomuv{e} \cr
r \, \rho_{uv} + r_u \rho_v + r_v \rho_u + 2 r \left( \phi_u \rho_v +
\phi_v \rho_u \right) = & \; 0 \; . & \eeomuv{f}
}
$$
It should be noted that when $\rho = 0$, the equations~\eeomuv{}
reduce to a real scalar field coupled to gravity.  This allows us to
check much of the code.

Since we shall be comparing results in different coordinate systems,
we need to know what invariant quantities can be built from the above
fields and metric. One quantity is the proper time $T$ on the $u=v$
axis, given by
\eqn\esuv{
T = \int a(u,u) \, du \; .
}
All curvature invariants can be constructed from the following three
building blocks. First, there is the Ricci scalar curvature
\eqn\eruv{
R = - {8 \over a^2} \left( \phi_u \phi_v + e^{4 \phi} \rho_u \rho_v
\right) \; ,
}
where we have used the equations of motion~\eeomuv{}. Two other
curvature invariants exist, defined by
\eqn\erauv{
R_2 \equiv  \sqrt{ {1 \over 32} \left( R_{ab} R^{ab} - R^2 \right)} =
{e^{2 \phi} \over a^2} \left(\phi_u \rho_v - \phi_v \rho_u \right) \;
,
}
\eqn\erbuv{
R_3 \equiv \sqrt{ {1 \over 12} C_{abcd} C^{abcd}} - {1 \over 6} R =
{1 \over r^2 a^2} \left(a^2 + 4 r_u r_v \right) = 2 m(u,v)/r^3 \; ,
}
where $C_{abcd}$ is the Weyl tensor, and $m(u,v)$ is the Hawking mass.
It is simple to show, e.g., using boundary conditions given later in
this section, that on axis, $u=v$, $R_2 = 0$, and $R_3 = -R/6$.

We can convert the system of equations~\eeomuv{} to a first order
system by defining new variables
\eqn\enew{
\rpsi = \phi_v\; , \;  \rpsib = \phi_u\; , \;
\rxi = \rho_v\; , \;  \rxib = \rho_u\; , \;
\rh = r_v\; , \;  \rhb = r_u\; , \;
\re = a_v/a \; .
}
In these new variables, \eeomuv{} becomes
\eqna\efirst
$$\eqalignno{
r \, \rh_u + \rh \rhb + {1 \over 4} a^2 = 0 \; , && \efirst{a} \cr
r \, \rhb_v + \rh \rhb + {1 \over 4} a^2 = 0 \; , && \efirst{b} \cr
\re_u - ( \rh \rhb + {1 \over 4} a^2 )/r^2 + \left( \rpsi \rpsib
+ e^{4 \phi} \rxi \rxib \right) = 0 \; , && \efirst{c} \cr
h_v - 2 e h + r \left( \rpsi^2 + e^{4 \phi} \rxi^2 \right) = 0 \; , &&
\efirst{d} \cr
r \, \rpsi_u + \rhb \rpsi + \rh \rpsib - 2 r e^{4\phi} \rxi \rxib = 0
\; , && \efirst{e} \cr
r \, \rpsib_v + \rhb \rpsi + \rh \rpsib - 2 r e^{4\phi} \rxi \rxib = 0
\; , && \efirst{f} \cr
r \, \rxi_u + \rhb \rxi + \rh \rxib + 2 r \left( \rpsib \rxi + \rpsi
\rxib \right) = 0 \; , && \efirst{g} \cr
r \, \rxib_v + \rhb \rxi + \rh \rxib + 2 r \left( \rpsib \rxi + \rpsi
\rxib \right) = 0 \; , && \efirst{h}
}$$
to which we adjoin
$$\eqalignno{
a_v - a e = 0 \; , && \efirst{i} \cr
r_v - h = 0 \; , && \efirst{j} \cr
\rho_v - \xi = 0 \; , && \efirst{k} \cr
\phi_v - \psi = 0 \; , && \efirst{l}
}$$
which follow from eq.~\enew.

Next we write down the boundary conditions which apply on axis,
$u=v$. Obviously, $r=0$ there, which implies $r_u + r_v = 0$, i.e.,
$\rhb = - \rh$. Eq.~\efirst{c} will be regular on axis if and only if
$a = 2 \rh$ (assuming $a>0$). Now~\efirst{g} implies $\rxi =
\rxib$ and~\efirst{f} implies $\rpsi = \rpsib$ on axis. From the
definitions~\enew, these imply $\rho_r = \phi_r = 0$ on axis. It
follows from a power series expansion (in powers of $r$) that $a_r =
0$ on axis.

Suppose that the fields $e$, $\rpsi$, and $\rxi$ are known as
functions of $v$ on a surface $u=u_0 = {\rm constant}$. We can
integrate~\efirst{l} as an ordinary differential equation in $v$ for
$\phi$. The initial data come from the boundary condition $\phi_r = 0$
on axis. Eq.~\efirst{k} for $\rho$ and~\efirst{i} for $a$ are treated
in identical fashion. Next, eqs.~\efirst{d},~\efirst{j} are integrated
as simultaneous equations for $r$ and $h$, with $r=0$, $h= a/2$ on
axis. Eq.~\efirst{b} with data $\rhb = -\rh$ on axis gives $\rh$ as a
function of $v$. Finally we integrate~\efirst{f},~\efirst{h} with data
$\rpsib = \rpsi$, $\rxib = \rxi$ on axis, to obtain $\rpsib$ and
$\rxib$ as functions of $v$. Thus we now have all quantities on the
surface $u=u_0$. Next we determine $e$, $\rpsi$, and $\rxi$ on the
surface $u=u_0 + \Delta u= {\rm constant}$ by integrating~\efirst{c},
\efirst{e} and \efirst{g}. Returning to the start of this paragraph,
we continue the integration process until either a singularity forms
or the fields have evaporated away leaving flat spacetime.

In order to start the integration we need to give initial data $e$,
$\rpsi$, and $\rxi$ on the initial surface $u=0$. The remaining
coordinate freedom is equivalent to the choice of $e$ (or $a$) on the
surface $u=0$. This is arbitrary. The initial fields $\rpsi$ and
$\rxi$ determine the physics.

The $u$-integration is done using an explicit finite difference
algorithm. The $v$-integration is done using an implicit (midpoint)
algorithm. However, if the equations are integrated in the manner
described above, they are linear and so the method is effectively
explicit. We operate exclusively at a Courant number of 1, and the
implicit algorithm provides the necessary stability. There is no need
for any artificial viscosity.

During the evolution fine structure appears on ever-decreasing
scales. We use the adaptive mesh algorithm of Berger and Oliger~\beol\
as modified in~\hast. This enables us to insert extra grid points
(again with a Courant number of 1) where and when necessary to resolve
the fine detail.

\newsec{The analytic critical solution}

It was conjectured in~\ehh\ that the precisely critical solution of
the axion/dilaton system coupled to gravity is continuously
self-similar. The continuously self-similar solution is described in
detail there, but we review it here. Continuous self-similarity means
that there exists a homothetic Killing vector $\zeta$ that satisfies
\eqn\ekill{
{\cal L}_\zeta g_{ab} = 2 g_{ab} \; .
}
The homothetic symmetry mixes with the $SL(2,\IR)$ symmetry, changing
$\tau$ by an infinitesimal $SL(2,\IR)$ transformation
\eqn\ektau{
{\cal L}_\zeta \tau = \alpha_0 + \alpha_1 \tau + \alpha_2 \tau^2 \; ,
}
with $\alpha_i \in \IR$.

The standard approach to spherically symmetric self-similar solutions
is to assume the metric has the form\footnote{*}{Ideally, one would
use double-null coordinates as in the preceding section. For unknown
technical reasons, we have been unable to repeat the calculations
of~\ehh\ in a $u-v$ coordinate system.}
\eqn\egrt{
ds^2 = (1 + u(t,r)) \left( -b^2(t,r) \, dt^2 + dr^2\right) + r^2 \,
d\Omega^2 \; .
}
The time coordinate $t$ is chosen so that the singular point occurs at
$t=0$, and the metric is regular for $t<0$. The homothetic vector
$\zeta$ is $\zeta = t \partial_t + r\partial_r$. One then defines a
new scale invariant coordinate $z \equiv -r/t$. In terms of the $z-t$
coordinates, the metric is
\eqn\egzt{
ds^2 = (1 + u(z)) \left[ \left(z^2 - b^2(z)\right) \, dt^2 + 2 z t \,
dt \, dz + t^2 \, dz^2 \right] + z^2 t^2 \, d\Omega^2 \; ,
}
and $\zeta$ becomes simply $\zeta = t \partial_t$. After fixing
unimportant constants, the axion/dilaton field has the form
\eqn\etzt{
\tau(t,z) = i { 1 - |t|^{i \omega} f(z) \over 1 + |t|^{i \omega} f(z)
} \; ,
}
where $\omega$ is a real constant, and $f(z)$ is a complex
function. We will sometimes write $f(z) = f_m(z) e^{i f_a(z)}$, where
$f_m$ and $f_a$ are real functions.

Details of the construction of the solution can be found in~\ehh. The
$r=0$ axis corresponds to $z=0$ and is a singular point of the
equations of motion. One can choose $b(z=0) = 1$ for $t<0$. The past
light cone of the origin corresponds to the line where $b(\zp) = \zp$,
and is also a singular point of the equations of motion. Requiring
regularity along the $r=0$ axis (for $t<0$) and regularity at $\zp$
uniquely fixes the solution in terms of $\omega$, $\zp$, $|f(0)|$, and
$|f(\zp)|$, with
\eqna\eztans
$$\eqalignno{
\omega = &\;  1.17695272200 \pm 0.00000000270 \; , & \eztans{a} \cr
\zp = &\;  2.60909347510 \pm 0.00000000216 \; , & \eztans{b} \cr
|f(0)| = &\; 0.892555411872 \pm 0.000000000224 \; , & \eztans{c} \cr
|f(\zp)| = &\; 0.364210875022 \pm 0.000000000760 \; . & \eztans{d}
}
$$

We can now construct the same invariant quantities as in the $u-v$
coordinates. On the $r=0$ axis, $u(z=0) = 0$ and $b(0) = 1$, so the
proper time is simply $T=t$. The curvature invariants are
\eqn\escalzt{
R = {2 \over t^2 z (1 + u(z))} \left[ {2 \omega |f(z)|^2 f_a'(z) \over
(1 - |f(z)|^2)^2} - {u(z) \over z} \right]
\; , }
\eqn\erazt{
R_2 = {\omega |f(z)| f_m'(z) \over 2 t^2 b(z) (1 + u(z)) (1 -
|f(z)|^2)^2} \; ,
}
\eqn\erbzt{
R_3 = { u(z) \over t^2 z^2 (1 + u(z))} \; .
}
We can use regularity on the $r=0$ axis to determine that
\eqn\escalzta{
R(z=0) = - { 2 \omega^2 |f(0)|^2 \over T^2 (1 - |f(0)|^2)^2} = -
{53.38 \over T^2} \; ,
}
and $R_2(z=0) = 0$, $R_3(z=0) = -R/6$, as expected. On the null cone
$b(\zp) = \zp$, there is no proper time. Define $\Lambda = \left(1 +
(\omega^2 - 2) |f(\zp)|^2 + |f(\zp)|^4\right)$. Using regularity, we
find
\eqna\escalztb
$$\eqalignno{
R(\zp) = &\;  - {2 \omega^2 |f(\zp)|^2 (1 - |f(\zp)|^2)^2 \over
t^2 \zp^2
\Lambda
\left(1 + (\omega^2 - 2) |f(\zp)|^4 + |f(\zp)|^8\right)} \; , &
\escalztb{a} \cr
R_2(\zp) = & \; - {\omega |f(\zp)|^2 (1 - |f(\zp)|^2)^3 (1 +
|f(\zp)|^2) \over
4 t^2 \zp^2
\Lambda
(1 + (\omega^2 - 2) |f(\zp)|^4 + |f(\zp)|^8)} \; , & \escalztb{b} \cr
R_3(\zp) = &\; {\omega^2 |f(\zp)|^2 \over
t^2 \zp^2
\Lambda
} \; . & \escalztb{c}
}$$
To compare with the numerical solution, we will find it more
convenient to consider just the ratios, or
\eqna\erats
$$\eqalignno{
{R_2(\zp) \over R(\zp)} = &\; {1 - |f(\zp)|^4 \over 8 \omega
|f(\zp)|^2} = 0.7866 \; , & \erats{a} \cr
{R_3(\zp) \over R(\zp)} = &\; - {1 + (\omega^2 - 2) |f(\zp)|^4 +
|f(\zp)|^8 \over 2 |f(\zp)|^2 (1 - |f(\zp)|^2)^2} = -4.958 \; , &
\erats{b} \cr
{R_2(\zp) \over R_3(\zp)} = &\; - {
(1 - |f(\zp)|^2)^3 (1 + |f(\zp)|^2)
\over
4 \omega (1 + (\omega^2 - 2) |f(\zp)|^4 + |f(\zp)|^8)
} = -0.1587 \; . &
\erats{c}
}$$

\newsec{Comparison of approaches at the critical solution}

As in~\hast, we can tune the initial conditions to be very close to
the threshold of black hole formation. This should, in principle,
result in a solution very close to the exactly critical solution of
the previous section. This behavior is most clearly displayed on the
$r=0$ axis, where we expect the numerical solution to resemble most
closely the exactly critical solution.

\ifig\onaxis{The scalar curvature as a function of proper time on
axis. The solid line is a slightly sub-critical numerical
solution. The dashed line is a slightly super-critical numerical
solution. The dotted line is the theoretical prediction~\escalzta.}{
\epsfxsize= .80\hsize
\epsfbox{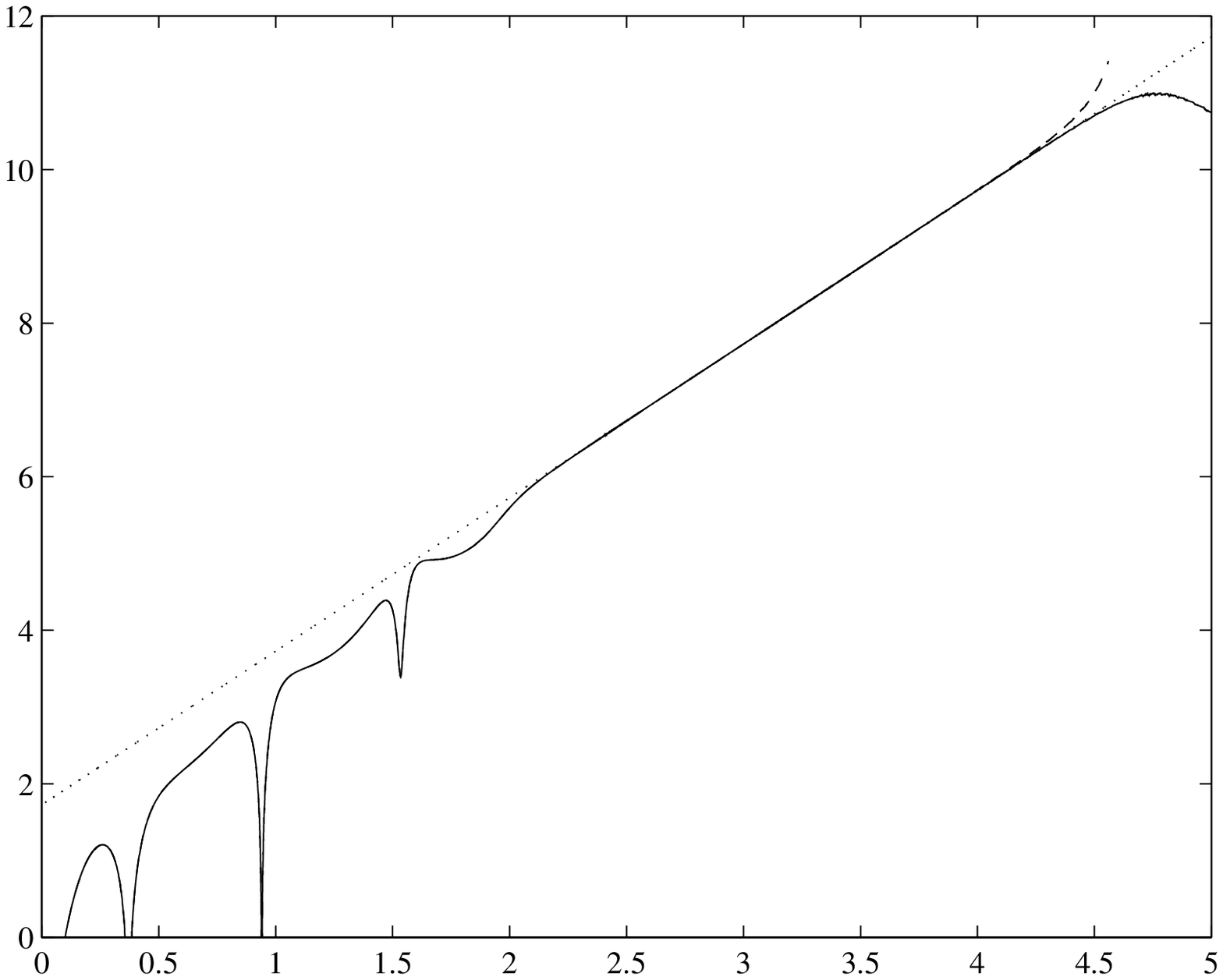}
\ifx\answ\bigans
\beginlabels\refpos 506 335 {}
\put 277 325 {- \log( T_{\rm crit} - T)}
\put 90 505 {\log( -R )}
\endlabels
\else
\beginlabels\refpos 506 335 {}
\put 327 325 {- \log( T_{\rm crit} - T)}
\put 195 455 {\log( -R )}
\endlabels
\fi
}

The scalar curvature on axis as a function of proper time is shown in
\onaxis. The initial conditions chosen for the numerical solution
shown contain much more dilaton than axion, but the late time behavior
is independent of initial conditions. The $x$-axis is $- \log(T_{\rm
crit} - T)$ (all logarithms in this paper have base~10), so the system
is evolving from left to right. For the analytical solution, $T_{\rm
crit} = 0$. For the numerical solutions, we are free to adjust $T_{\rm
crit}$ to most closely match the analytical solution. The $y$-axis is
$\log(-R)$. Three lines are shown in \onaxis. The solid line is a
slightly sub-critical numerical evolution. The curvature becomes
strong on axis, but eventually dissipates (when $T > T_{\rm
crit}$). The dashed line is a slightly super-critical solution. It
becomes a black hole, as can be seen by its upturn for late $T$. The
dotted line is the continuously self-similar solution~\escalzta.

As expected, on the left side of \onaxis, the numerical and analytical
solutions disagree. This is the region where the numerical solutions
depends on the initial values of the fields. As the evolution
progresses, the numerical solutions approach the analytical solution,
becoming almost indistinguishable for a while.  This is strong
confirmation that the continuously self-similar solution is indeed the
attractor.  Eventually, the numerical solutions diverge from each
other and from the analytical solution.

\ifig\ratba{The ratio $R_2/R$ as a function of $- \log r$
along a line of constant
$v$. The solid line is a slightly subcritical numerical solution. The
dotted line in the theoretical prediction~\erats{a}. As $r \rightarrow
0$, $R_2/R \rightarrow 0$.}{
\epsfxsize= .8\hsize
\epsfbox{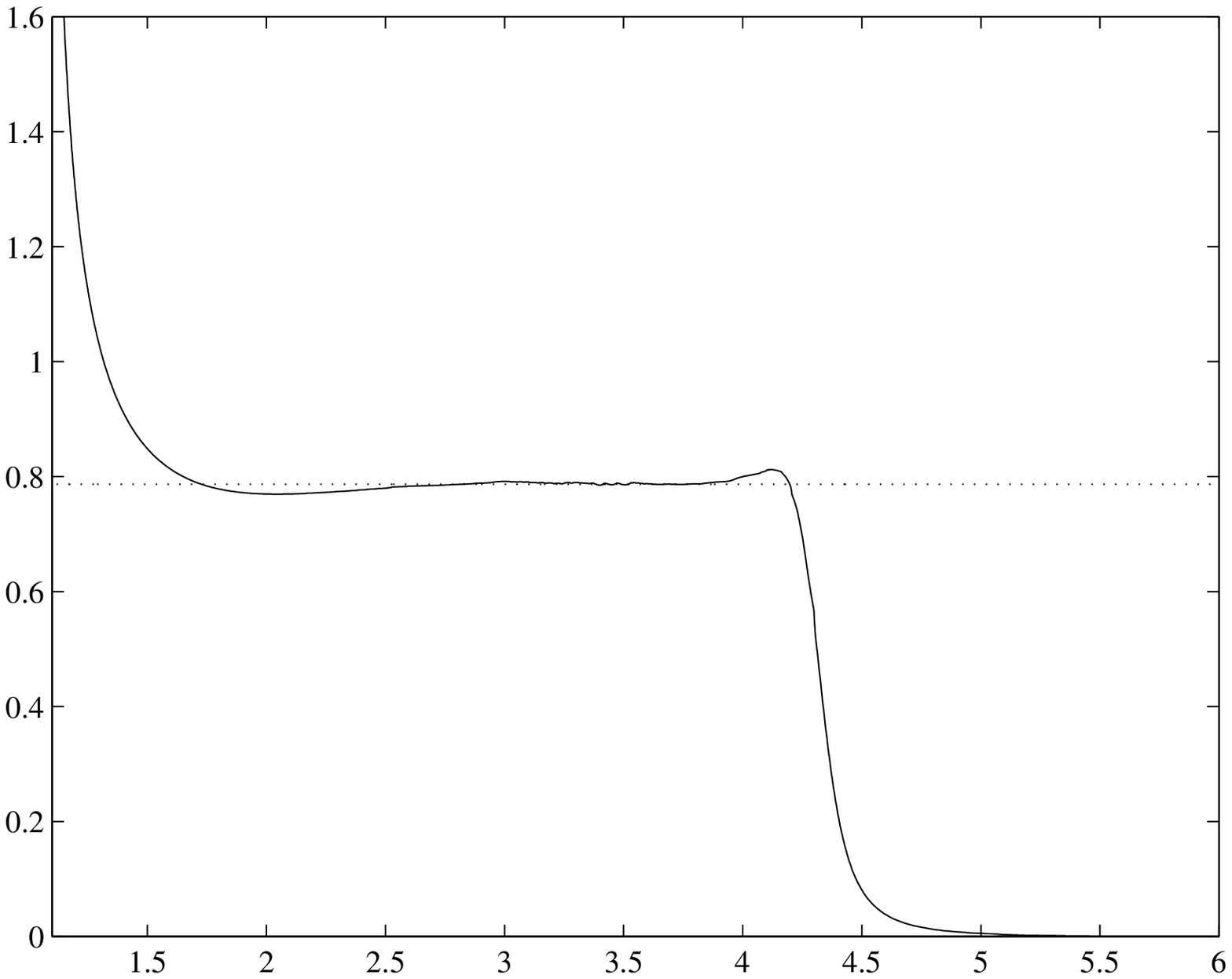}
\ifx\answ\bigans
\beginlabels\refpos 495 185 {}
\put 327 174 {-\log r}
\put 95 335 {{R_2 \over R}}
\endlabels
\else
\beginlabels\refpos 495 185 {}
\put 347 174 {-\log r}
\put 205 335 {{R_2 \over R}}
\endlabels
\fi
}
\ifig\ratca{The ratio $R_3/R$ as a function of $- \log r$
along a line of constant
$v$. The solid line is a slightly subcritical numerical solution. The
dotted line in the theoretical prediction~\erats{b}. As $r \rightarrow
0$, $R_3/R \rightarrow -R/6$.}{
\epsfxsize= .8\hsize
\epsfbox{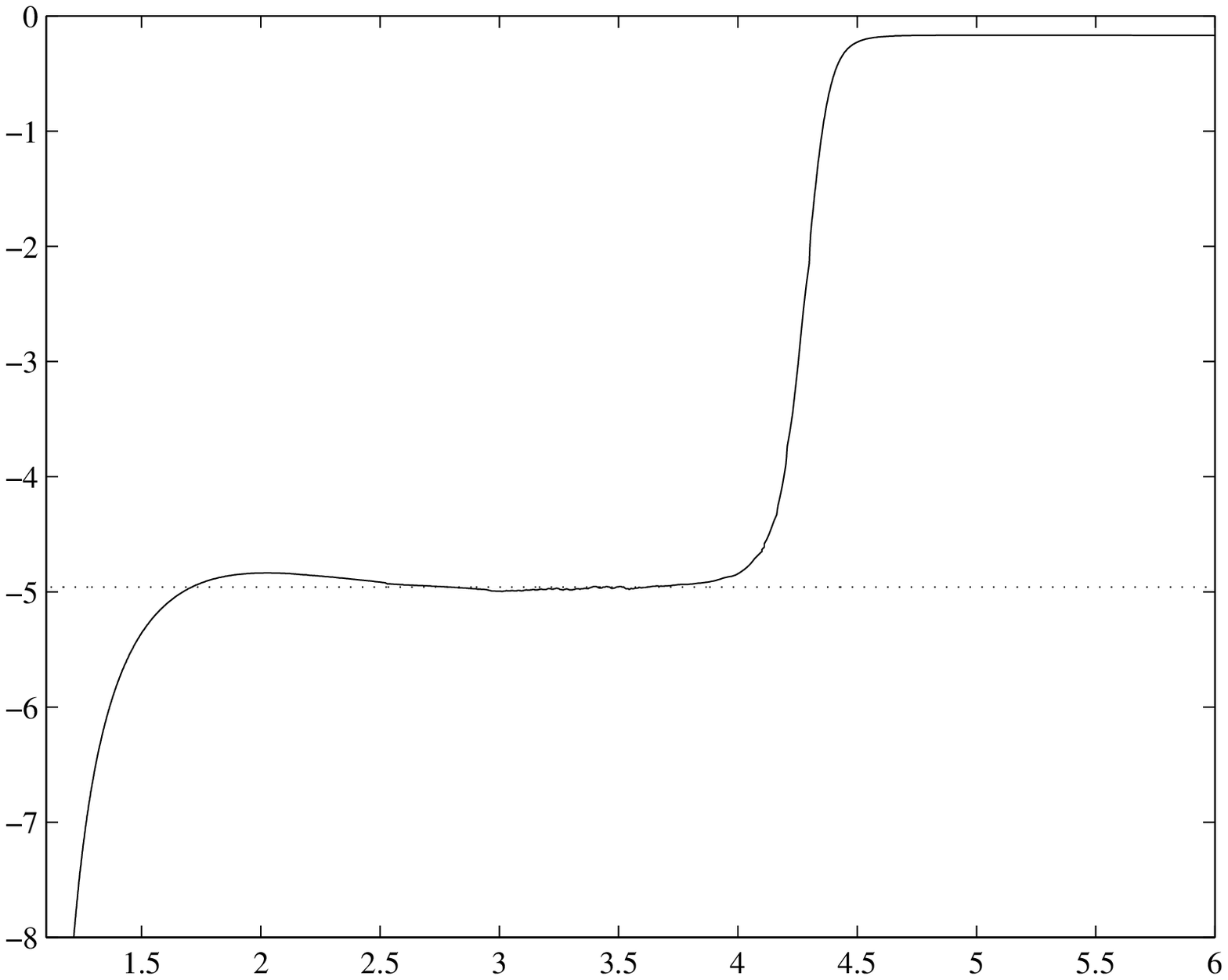}
\ifx\answ\bigans
\beginlabels\refpos 495 185 {}
\put 327 174 {-\log r}
\put 95 335 {{R_3 \over R}}
\endlabels
\else
\beginlabels\refpos 495 185 {}
\put 347 174 {-\log r}
\put 205 335 {{R_3 \over R}}
\endlabels
\fi
}
\ifig\ratbc{The ratio $R_2/R_3$ as a function of $- \log r$
along a line of constant
$v$. The solid line is a slightly subcritical numerical solution. The
dotted line in the theoretical prediction~\erats{c}. As $r \rightarrow
0$, $R_2/R_3 \rightarrow 0$.}{
\epsfxsize= .8\hsize
\epsfbox{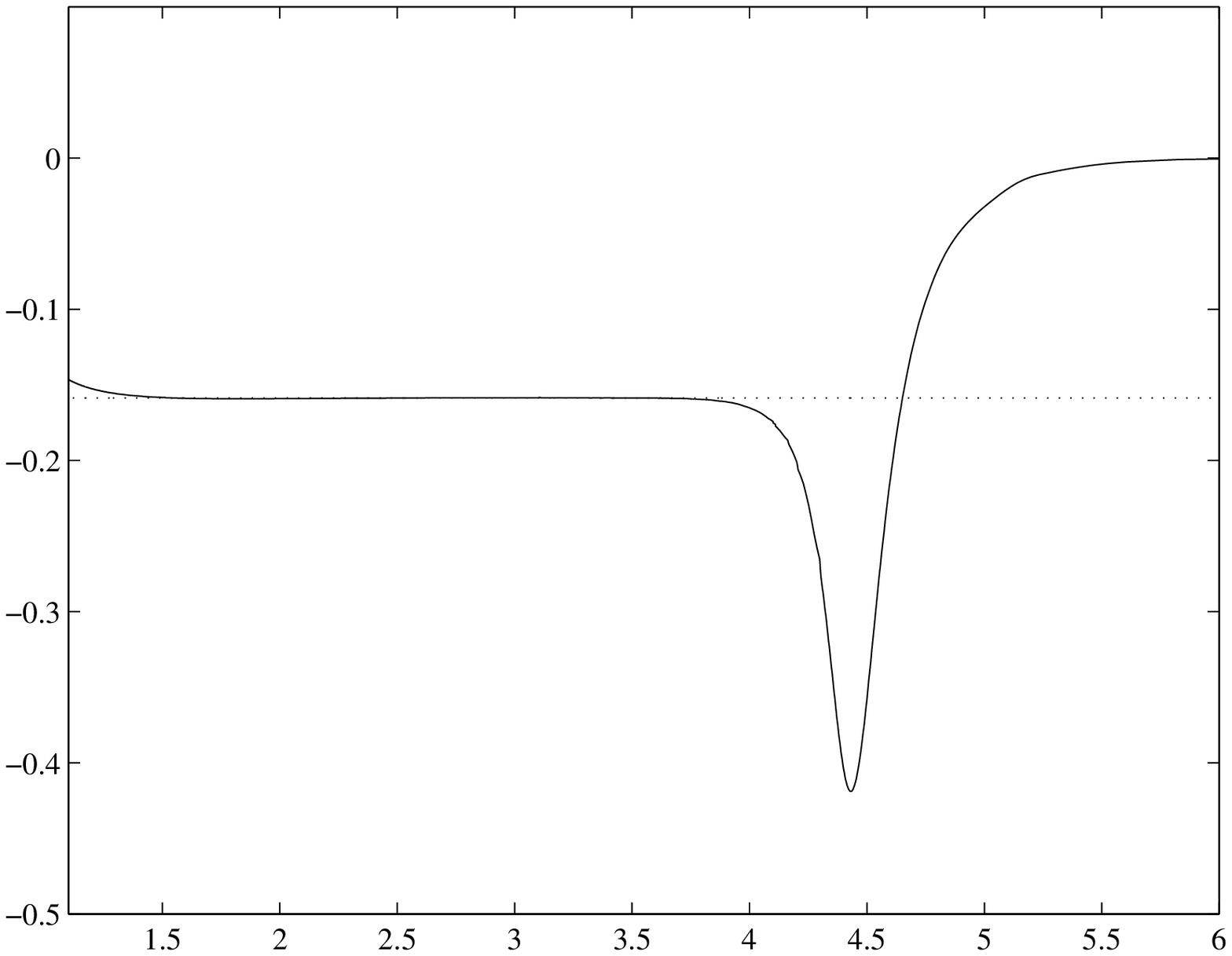}
\ifx\answ\bigans
\beginlabels\refpos 495 185 {}
\put 327 174 {-\log r}
\put 95 335 {{R_2 \over R_3}}
\endlabels
\else
\beginlabels\refpos 495 185 {}
\put 347 174 {-\log r}
\put 205 335 {{R_2 \over R_3}}
\endlabels
\fi
}

We compare the solutions on the past lightcone of the singular point
in \ratba, \ratca\ and \ratbc.  From the analytic point of view, this
is straightforward. Eq.~\erats{} tells us that the ratios of the
curvature invariants should approach constants.  Numerically, it is
much trickier. The past lightcone corresponds to a line of constant $v
= v_{\rm crit}$. One must choose $v_{\rm crit}$ carefully because any
error will grow as it approaches the $r=0$ axis. Unfortunately, this
is precisely where the numerical solution should be approaching the
analytical solution. To further complicate matters, since the
lightcone is a null curve, there is no invariant definition of
distance along the curve. In the figures, we have plotted the
curvature ratios as a function of $-\log r$. The function $r$ becomes
multivalued for supercritical evolutions (which is why $r$ is a bad
choice as a coordinate), so we show a slightly subcritical numerical
evolution in the figures.  We see that the numerical solution (shown
as a solid line) approaches the analytical line (the dotted line),
remains there for a while, and then diverges from the analytical
solution.  When $r=0$, we expect $R_2 = 0$ and $R_3 = -R/6$, and these
are indeed the values seen for very small $r$. The transition from the
$z=\zp$ values~\erats{} to the $r=0$ values is fairly abrupt. The
transition is at finite $r$ because of errors in choosing $v_{\rm
crit}$ and because the numerical solution is not precisely
critical. Note, for example, that $R_3(\log r = -4) \approx 10^7$ and
is growing quickly near $\log r = -4$.  Thus the fact that the ratios
are roughly constant and close to the analytical prediction~\erats{}
is an excellent indication that the solution is indeed the
continuously self-similar solution given in~\S{4}.

\newsec{Moving away from criticality}

One of the most interesting findings in~\chop\ was the discovery of
the scaling law~\escal\ for the mass of black holes away from
criticality.  The striking result was that $\gamma$ was completely
independent of initial conditions, and was thus dubbed the
``universal'' critical exponent. For the case of a real scalar field
coupled to gravity, numerical work has given $\gamma =
0.374$~\refs{\chop,\hast}.  Previous numerical work with other types
of matter gives similar values for $\gamma$. A study of axisymmetric
gravitational wave collapse~\abev\ indicates $\gamma \approx
0.36$. Spherically symmetric perfect radiation fluid~\evco\ gives
$\gamma \approx 0.36$. This coincidence of numbers led people to
suspect that $\gamma$ might be a universal exponent, independent of
the specific type of matter coupled to gravity.

The universality of $\gamma$ has been drawn into question by
analytical arguments. Starting with a known critical solution, one can
use renormalization group techniques to calculate $\gamma$. This was
first carried out for the perfect radiation fluid in~\kha, which found
$\gamma = 0.3558$, in good agreement with~\evco, but only fairly close
to the value for a real scalar field. Because of the numerical
uncertainties in~\evco\ and uncertainties in the accuracy of the
analytic arguments, it is unclear whether these results indicate
conclusively that $\gamma$ depends on the matter content. A more
general class of perfect fluids with $p = k \rho$ was studied
analytically in~\mai\ under the assumption that the critical solution
is continuously self-similar. It was found that $\gamma$ depends
strongly on $k$. However, no numerical work has been done for $k\ne
1/3$, and it is not clear that the continuously self-similar solution
really is the attractor at criticality.

The case of a complex scalar field has been studied in~\hieab. They
also perturbed away from a continuously self-similar solution, and
found $\gamma = 0.3871$, which seems to be distinctly different from
the perfect radiation fluid value. However, they also found that the
continuously self-similar solution is unstable, and conjecture that it
is an attractor of codimension three. The discretely self-similar
solution (the same critical solution as for the real scalar field
case) has codimension one, and therefore dominates. Thus numerically
it should be difficult to find the continuously self-similar
solution. We have numerically evolved the complex scalar field with a
variety of initial conditions and in every case have found critical
behavior of the discretely self-similar type.  Thus, no conclusive
evidence has so far been presented that $\gamma$ is not universal.

Given an exactly critical analytical solution, one can perturb the
critical solution to find the critical exponent $\gamma$ as
follows~\kha. Let $h$ be any function of the analytical solution, such
as $b$ or $f$. Perturb away from the critical solution
\eqn\ehdef{
h_t(z,t) = h_{\rm ss}(z) + \epsilon |t|^{- \kappa} h_{\rm pert}(z) \; ,
}
where $h_{\rm ss}(z)$ is the critical solution, $\epsilon$ is a small
number, $\kappa$ is a constant, and $h_{\rm pert}(z)$ depends only on
$z$. Replacing $h_t(z,t)$ in the equations of motion and keeping terms
to first order in $\epsilon$ gives an eigenvalue equation for
$\kappa$. This eigenvalue equation can in principle have a number of
possible solutions for $\kappa$. The solution with the largest value
of $\Re \kappa$ will cause the fastest growing perturbation in~\ehdef,
and is called the ``most relevant mode.'' The critical exponent is
given by~\refs{\kha,\mai,\hieab}
\eqn\ecrit{
\gamma = {1 \over \Re \kappa} \; .
}

We have carried out a perturbation analysis of the axion/dilaton
critical solution described in \S{4}. In principle, $\kappa$ can have
an imaginary component, but this greatly complicates the analysis, and
we found it to be unnecessary in this instance. The result for
perturbation of the continuously self-similar solution is
\eqn\ecrita{
\gamma_{\rm analytic} = 0.2641066 \; .
}
This value is quite different from the critical exponents found in
previous numerical studies, and therefore should be easy to
distinguish numerically. Our analysis does not rule out other possible
values of $\kappa$, but we have seen no numerical evidence for other
unstable modes.

\ifig\mass{Mass as a function of $p - p_{\rm crit}$ for the
axion/dilaton system.}{
\epsfxsize= 0.8\hsize
\epsfbox{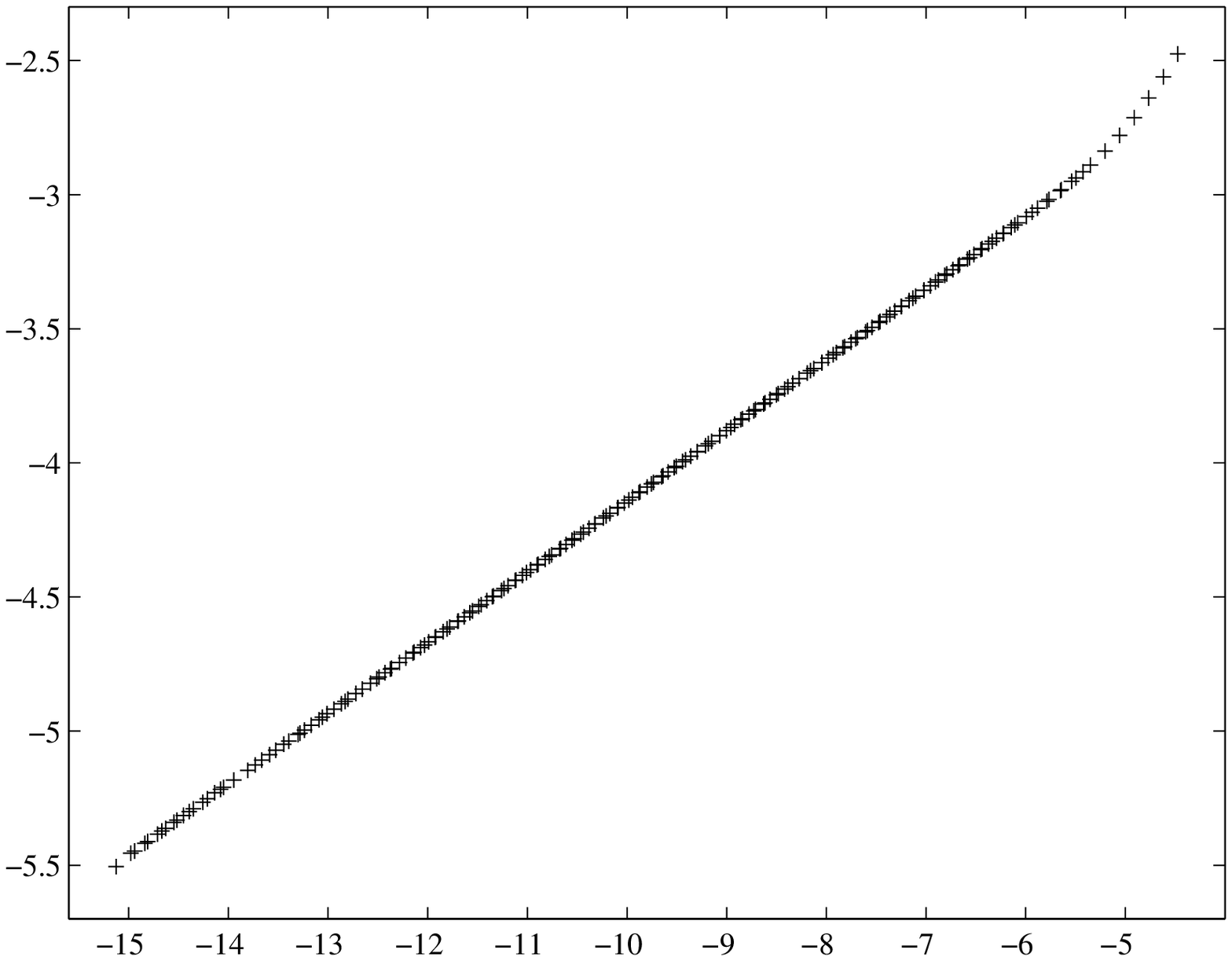}
\ifx\answ\bigans
\beginlabels\refpos 495 477 {}
\put 275 460 {\log( p - p_{\rm crit})}
\put 75 640 {\log M_{\rm bh}}
\endlabels
\else
\beginlabels\refpos 495 477 {}
\put 335 464 {\log( p - p_{\rm crit})}
\put 190 640 {\log M_{\rm bh}}
\endlabels
\fi
}

We show in \mass\ the numerical results for a large number of
supercritical evolutions along a one-parameter family of initial
conditions parameterized by $p$. The mass is defined as in~\hast. Pick
a constant $v_0$ and determine the point $(u_0, v_0)$ where the
apparent horizon intersects $v_0$. The mass of the black hole is
defined as $M_{\rm bh} = m(u_0,v_0)$, where $m(u,v)$ is the Hawking
mass defined in~\erbuv. As can be seen from \mass, $\log M_{\rm bh}$
is a linear function of $\log (p - p_{\rm crit})$ until one is far
away from the critical solution. Omitting the final seven points in
\mass\ where the curve is no longer linear, we obtain a slope for
\mass\ of
\eqn\emsb{
\gamma_{\rm numeric} = 0.264 \; ,}
which agrees nearly perfectly with the theoretical
prediction~\ecrita. As in the case of the real scalar field,
$\gamma_{\rm numeric}$ seems to be independent of the initial data, as
long as the axion in not identically zero. This is further strong
evidence that the critical solution is indeed continuously
self-similar, and that the perturbation expansion gives the right
answer.

\newsec{Conclusions and unanswered questions}

\ifig\crit{Points along the critical line in the axion/dilaton
system. The $x$-axis measures the strength of the axion initial data,
and the $y$-axis measures the strength of the dilaton initial data.}{
\epsfxsize= 0.8\hsize
\epsfbox{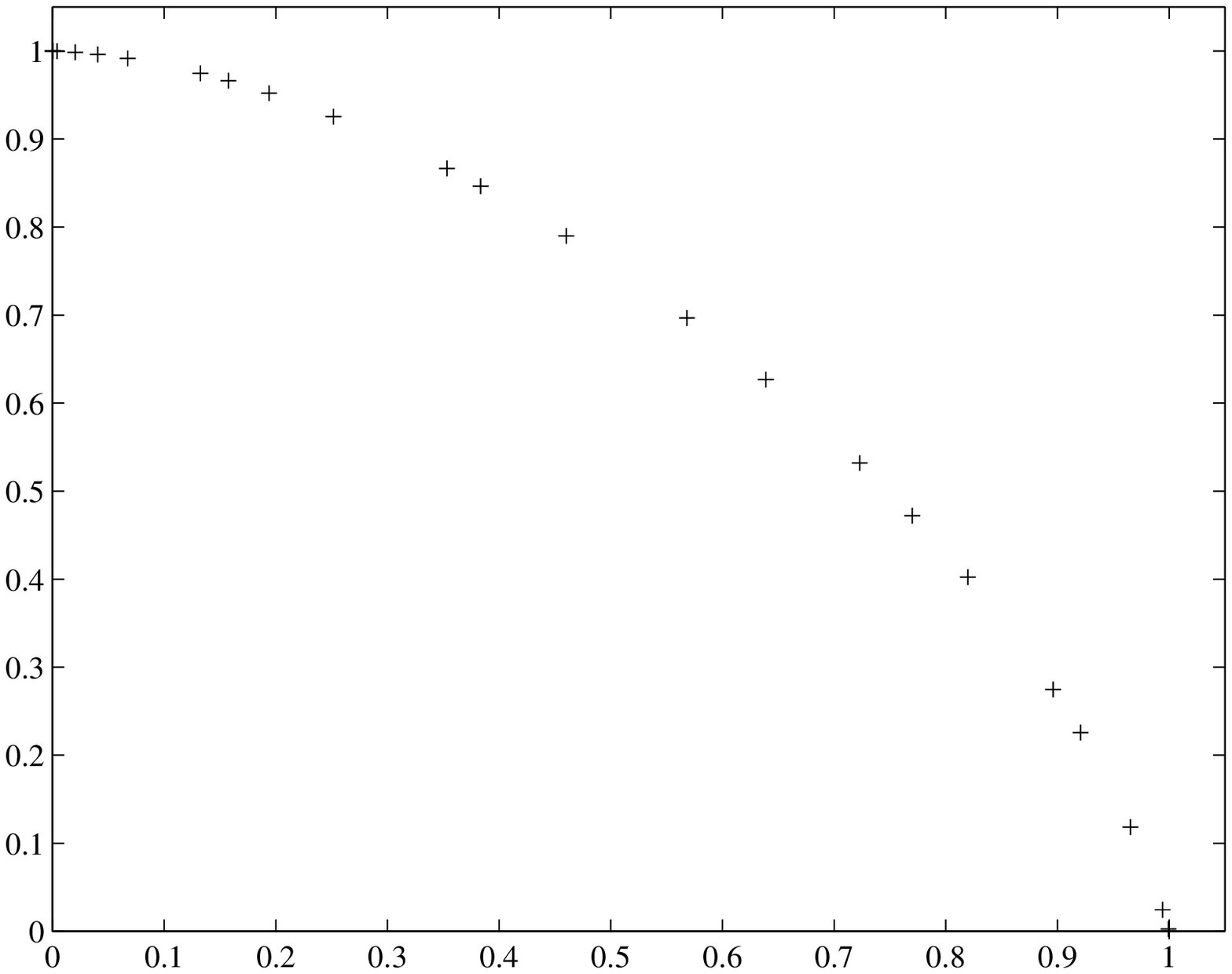}
\ifx\answ\bigans
\beginlabels\refpos 495 453 {}
\put 320 447 {\rho}
\put 112 634 {\phi}
\endlabels
\else
\beginlabels\refpos 495 453 {}
\put 370 442 {\rho}
\put 210 624 {\phi}
\endlabels
\fi
}

We depict a two-dimensional subspace of the infinite-dimensional space
of initial conditions in \crit. The $x$-axis is the (normalized)
amplitude of the initial axion field, and the $y$-axis is the
(normalized) amplitude of the initial dilaton field. The points on the
graph represent critical solutions. Not surprisingly, the critical
surface has codimension one. When the axion field is initially
identically zero, it remains zero, and the problem reduces to a real
scalar field coupled to gravity. The critical value for that solution
is the point on the $y$-axis, and we know from~\chop\ that the
critical solution there is discretely self-similar. The conjecture
of~\ehh\ is that the points on the critical line with nonzero axion
are continuously self-similar.  This raises the question: is the
discretely self-similar solution simply an artifact of special initial
conditions? In other words, is the discretely self-similar solution an
isolated point on the critical line, or is there an open set on the
critical line near the $y$-axis where the critical solution is
discretely self-similar? This is a difficult question to answer
numerically, because near the $y$-axis, the axion is quite small and
is thus liable to be swamped by numerical noise. The solution in
\onaxis\ represents a point quite near the $y$-axis, and is clearly
continuously self-similar. By moving along the critical line even
closer to the $y$-axis, we push the region in \onaxis\ that depends on
initial conditions farther to the right. Eventually, this will meet
the point where the sub- and super-critical solutions diverge from the
analytical solution. Thus we cannot expect to move arbitrarily close
to the $y$-axis in a numerical evolution. Except for points at which
we believe the numerical accuracy is breaking down, every point near
the $y$-axis that we have checked is continuously self-similar. This
implies that the discretely self-similar solution is unstable, but
analytical work is necessary to confirm this.

The axion/dilaton matter system described in this paper is physically
well motivated. However, if the discretely self-similar solution is
simply the result of non-generic initial conditions, might the
continuously self-similar solution also be the result of overly
restricted matter? For instance, string theory contains numerous gauge
fields in the low-energy effective action which we have omitted in
writing~\eaction. Including these fields might result in yet another
type of critical solution. If we hope to use critical solutions to
probe the small-scale structure of spacetime, we should in principle
use the most general solution allowed in our favorite theory of
quantum gravity.

\bigskip
\centerline{\bf Acknowledgements}

This work was done on workstations supplied by the EPSRC (formerly
SERC) Computational Science Initiative Research Grant GR/H57585. RSH
was supported by an EPSRC studentship and JHH by EPSRC grant GR/H34937
and PPARC grant GR/K29272.

\listrefs
\bye